\documentclass{aa}
\usepackage{graphics,latexsym,psfig}

\def\etal{{et\,al.}}
\def\Ni{\noindent}
\def\msun{M$_{\odot}$}

\def\amin{\ifmmode ^{\prime}\else$^{\prime}$\fi}
\def\asec{\ifmmode ^{\prime\prime}\else$^{\prime\prime}$\fi}
\def\farcs{\hbox{$.\!\!^{\prime\prime}$}}  

\newbox\grsign \setbox\grsign=\hbox{$>$}
\newdimen\grdimen \grdimen=\ht\grsign
\newbox\laxbox \newbox\gaxbox
\setbox\gaxbox=\hbox{\raise.5ex\hbox{$>$}\llap
     {\lower.5ex\hbox{$\sim$}}}\ht1=\grdimen\dp1=0pt
\setbox\laxbox=\hbox{\raise.5ex\hbox{$<$}\llap
     {\lower.5ex\hbox{$\sim$}}}\ht2=\grdimen\dp2=0pt
\def\gax{\mathrel{\copy\gaxbox}}
\def\lax{\mathrel{\copy\laxbox}}
\def\rx0537{RX J0537.7--7034}
\def\1e{1E 0035.4--7230}

\begin{document}

   \thesaurus{06         
              (02.01.2;  
               08.02.1:  
               08.02.3;  
               08.09.2:  
               13.25.5)} 

    \title{\rx0537: The shortest-period supersoft X-ray source
         \thanks{Based on observations (60.H-0682 and 62.H-0839) 
           obtained at the European Southern Observatory, La Silla, Chile.}}

   \author{J. Greiner\inst{1}, M. Orio\inst{2}, and R. Schwarz\inst{1}}

   \offprints{J. Greiner, jgreiner@aip.de}

   \institute{Astrophysical Institute
        Potsdam, An der Sternwarte 16, D-14482 Potsdam, Germany
      \and
        Osservatorio Astronomico di Torino, Strada Osservatorio 20,
         I-10125 Pino Torinese (TO), Italy}

   \date{Received 18 October 1999; accepted 14 December 1999}

   \maketitle 

   \begin{abstract}

We present new photometric and spectroscopic observations of the transient 
supersoft X-ray source RX J0537.7--7034 and find a periodicity of 
approximately 3.5 hrs. This establishes \rx0537\ as the supersoft 
X-ray source with the shortest orbital period. We furthermore derive
an inclination of the binary system of 45\degr\ $\lax i \lax$ 70\degr,
and the masses of the two binary components: 
$M_{accretor} = 0.6 \pm 0.2$ \msun, $M_{donor} = 0.35 \pm 0.02$ \msun.
This implies that the 
standard scenario of supersoft X-ray sources, in which the donor
is thought to be more massive than the accreting white dwarf to ensure high
mass transfer rates on a thermal timescale (van den Heuvel \etal\ 1992), 
is not applicable for this system.
We discuss alternative interpretations of this source
as a former nova in which the thermonuclear  flashes  have become mild 
and most accreted mass is  retained by the white dwarf (so-called SMC 13 
type systems), or as a self-sustained wind-driven system.

      \keywords{X-ray: stars -- accretion disks -- binaries: close --
                stars: individual: \rx0537, \1e\ $\equiv$ SMC 13
               }

   \end{abstract}

\section{Introduction}

Supersoft X-ray sources (SSS) are persistent or transient sources whose
radiation is almost  completely emitted in the energy band below 0.5 keV and
whose bolometric luminosity is 10$^{37-38}$ erg s$^{-1}$ (close
to the Eddington limit for a 1 M$_\odot$ star). After the 
{\it Einstein} discovery of CAL 83 and CAL 87 (Long \etal\ 1981, 
Crampton \etal\ 1987, Pakull \etal\ 1987, 1988) this class of objects 
was established by
many ROSAT discoveries (see Orio 1995, Greiner 1996 and references therein).
Several different observational facts suggest that
all these very luminous sources contain white dwarfs and that 
many of these are burning hydrogen in a shell, with a very thin and
hot atmosphere on top. The white dwarf might be single 
(hot PG 1159 stars or planetary nebulae nuclei, e.g. Wang
 1991, Motch et al. 1993) or in a binary system. The binary
supersoft X-ray sources also belong to different types with
orbital periods ranging from $\simeq$1 year for symbiotic stars down
to $\simeq$4 hours for SMC 13 type sources (Kahabka and Ergma 1997
and references therein). 
The subgroup with short orbital periods ($\leq$3 days) now encompasses
9 sources in the Magellanic Clouds and the Milky Way. The widely accepted
scenario for these so-called close-binary supersoft sources (CBSS)
is based on a companion more massive than the white dwarf, so that
mass transfer is unstable and can occur at the high rates which are required
to burn the hydrogen steadily on the white dwarf surface (van den Heuvel 
\etal\ 1992, Rappaport \etal\ 1994).
While a recent review can be found in  van den Heuvel and Kahabka (1997),
we remind here that supersoft X-ray sources are of great
 astrophysical interest also because of their possible nature as 
progenitors of type Ia supernovae  or leading to a  neutron star by 
accretion induced collapse.    
 
The discovery of the  supersoft X-ray source RX J0537.7--7034 was
first announced  by Orio \& \"Ogelman (1993). Later Orio \etal\ (1997) showed
 that the source was transient in X-rays and  appeared ``on''
 for $\simeq$ 1 year. Also, the optical identification
 with a blue variable star at $V \simeq 19.7$ mag was proposed.
The optical spectrum shows the \ion{He}{ii} $\lambda$4686
 line in emission and the two Balmer lines H$\gamma$ and H$\delta$ in
 absorption. These spectral features appear redshifted by
almost 5 \AA, as is expected for a LMC membership. 
Finally, Orio \etal\ (1997) also showed that the photometric
data for \rx0537\ might be consistent with a possible orbital period
of only $\simeq$ 2.5--3 hours. 

Here we present new
photometric and spectroscopic  data of \rx0537\ that indeed prove
it to have a very short period. 

\begin{table*}
\caption{Log of all observations of \rx0537.}
\vspace{-0.2cm}
\begin{tabular}{lccccc} 
\hline \noalign{\smallskip}
Telescope/ & Date &  Spectral range  & No. of     & Duration &  $T_{\rm int}$\\
Instrument &      &                  &  exposures & (h)      &  (sec) \\
\noalign{\smallskip} \hline \noalign{\smallskip}
DFOSC 1.5~m & 1998 Jan 18 & $V/B$             & 19/22 & 3.8 & 120/180 \\
DFOSC 1.5~m & 1998 Jan 19 & $V$               & 61    & 2.6 & 120 \\
EFOSC 3.6~m & 1999 Jan 22 & 3800--8000 \AA  & 7     & 4.2 & 1800~\ \\
DFOSC 1.5~m & 1999 Jan 24 & $V$               & 74    & 4.0 & 120 \\
DFOSC 1.5~m & 1999 Jan 25 & $V$               & 81    & 4.5 & 120 \\
DFOSC 1.5~m & 1999 Jan 26 & $V$               & 74    & 2.5 & 120 \\
  \hline \noalign{\smallskip} 
  \end{tabular}

  \label{logtab}
  \end{table*}

   \begin{figure}
      \vbox{\psfig{figure=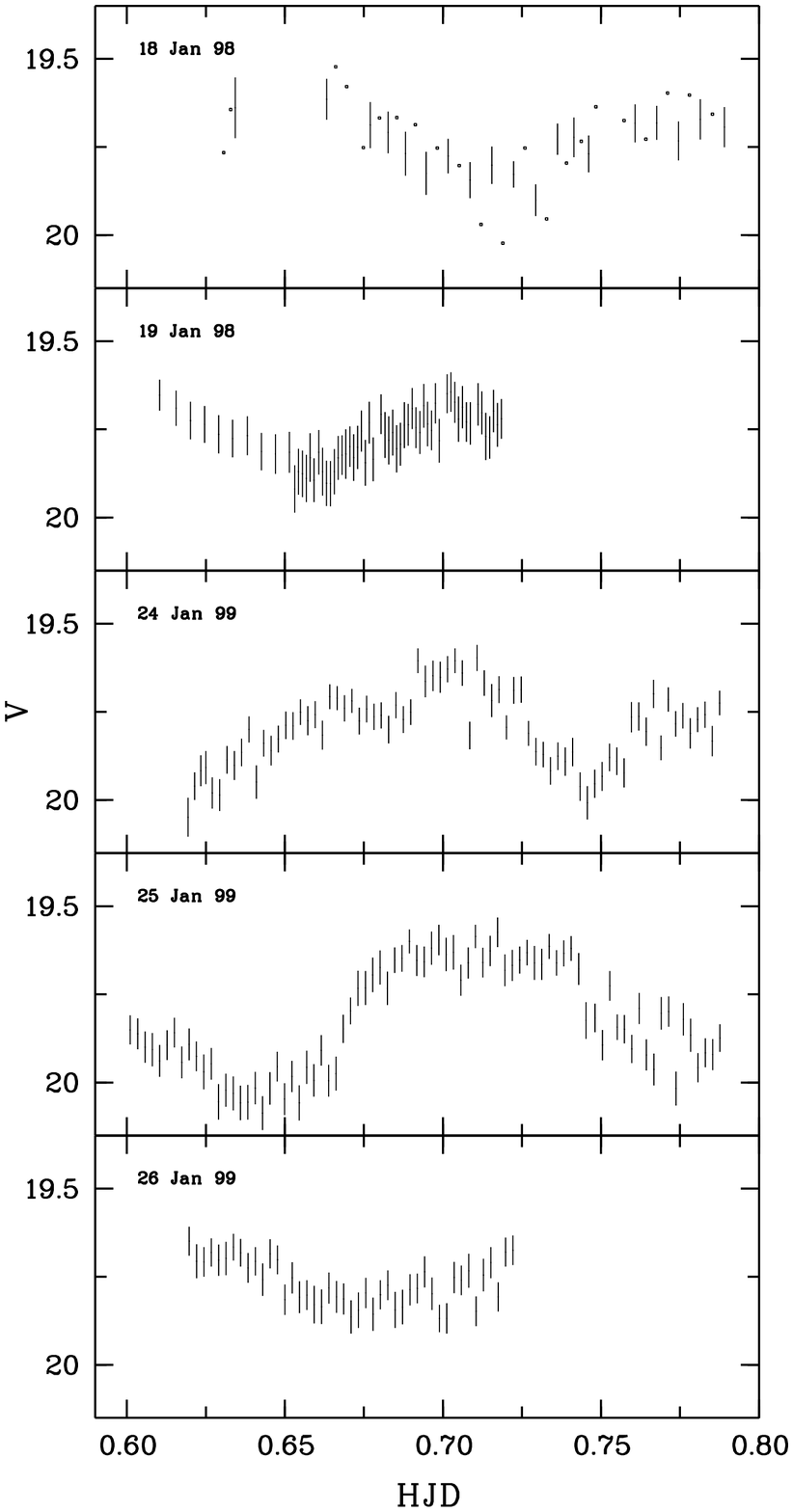,width=8.5cm,height=17.cm,%
           bbllx=2.3cm,bblly=1.2cm,bburx=15.8cm,bbury=26.5cm,clip=}}
    \caption[lc]{$V$ band light curves of \rx0537\ obtained during our 
         5 nights. 
        The top panel includes in addition the $B$ band data (filled squares)
        and the zero point of the x-axis is 0.05. 
           Note the strong variability from night to night.
             }
      \label{lc}
   \end{figure}

\section{Optical photometry}

We carried out new photometric observations of \rx0537\ at La Silla 
on January 18/19 1998 and January 24-26 1999
with the 1.5\,m Danish telescope of the European Southern Observatory
(ESO) equipped with the spectrograph-imager DFOSC.  
Observations were primarily done in the $V$ band, with one run also in $B$.
Exposure times typically were 2-3 min. each, and our total temporal coverage
amounts to more than 17 hours (see details in Tab. 1).
Photometric reduction in this very crowded field was done using the 
profile-fitting algorithm of {\sc DoPhot} (Mateo \& Schechter 1989).
On January 24 and 25, 1999 a set of Landolt (1992) standards were observed 
confirming the absolute photometric calibration derived by Orio \etal\ (1997).

   \begin{figure}
      \vbox{\psfig{figure=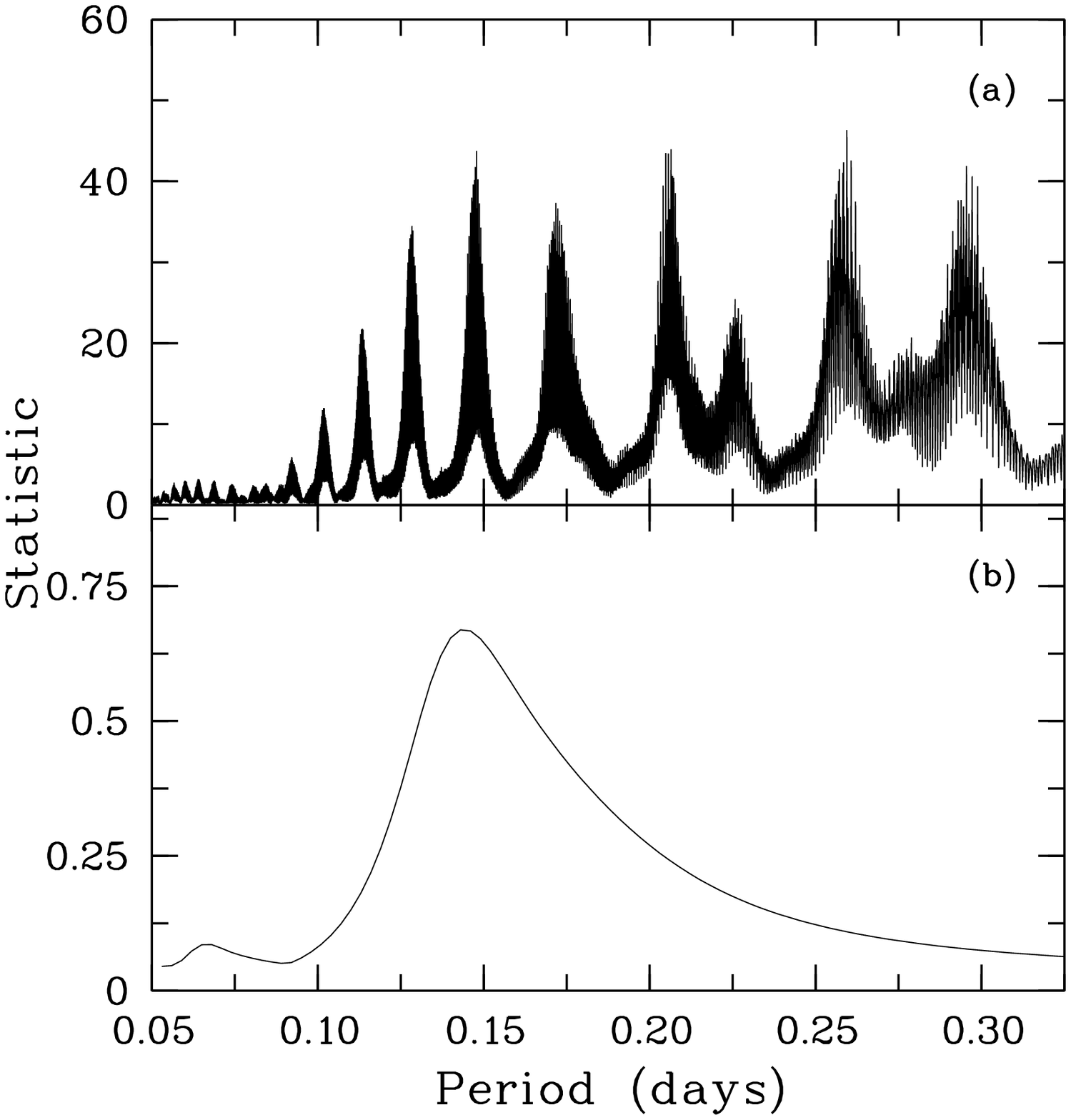,width=8.8cm,%
           bbllx=1.4cm,bblly=5.5cm,bburx=18.4cm,bbury=23.cm,clip=}}
    \caption[period]{Results of the period-search using the 
        analysis-of-variance method for the photometry (top)
        and radial velocities (bottom). The adopted orbital period
        is that peak in the top panel which best coincides with the
        spectroscopic peak (0.147275 d).}
      \label{perio}
   \end{figure}

   \begin{figure*}
      \vbox{\psfig{figure=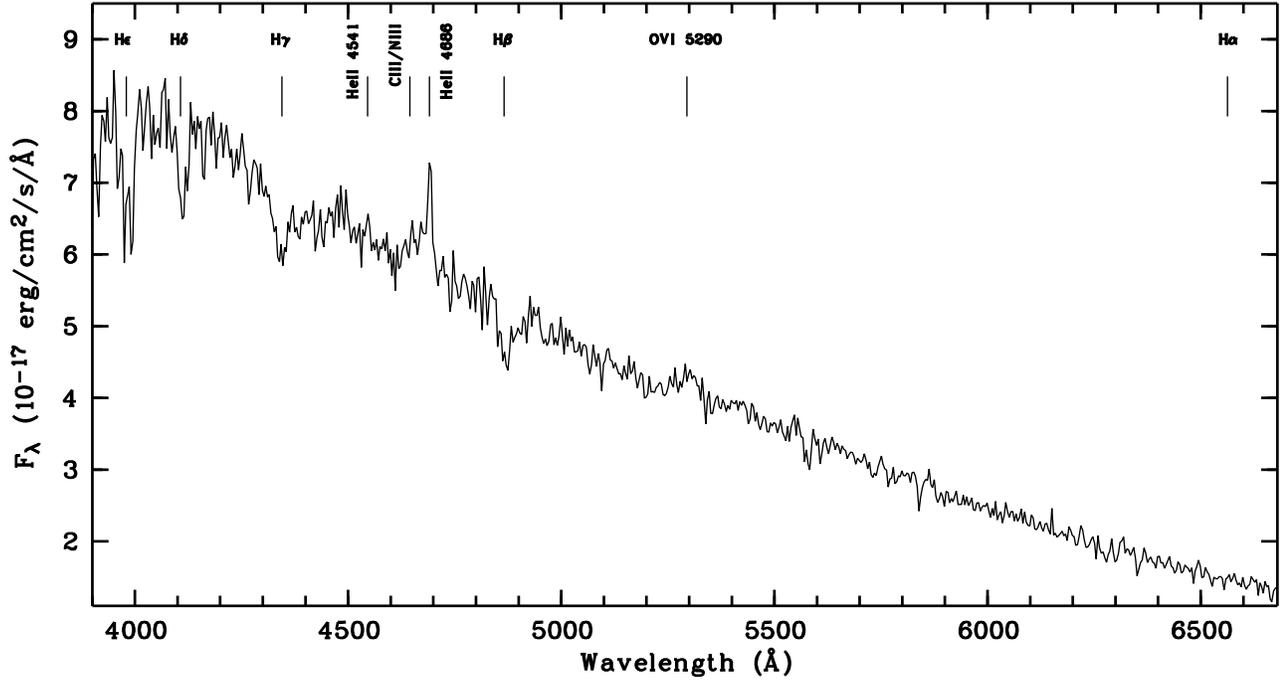,width=17.5cm,%
           bbllx=3.2cm,bblly=3.3cm,bburx=19.3cm,bbury=12.3cm,clip=}}
    \vspace*{-0.3cm}
    \caption[osp]{Sum of all 7 spectra corresponding to an exposure 
           time of 13\,300 sec. The \ion{He}{ii} $\lambda$4686 emission
           line is prominent, and other detected lines are also marked.
           Note that the Balmer series is in absorption. With the present
           signal-to-noise ratio the existence of possible emission cores
           within the absorption troughs remains unclear.
             }
      \label{ospec}
   \end{figure*}

   \begin{figure*}
      \vbox{\psfig{figure=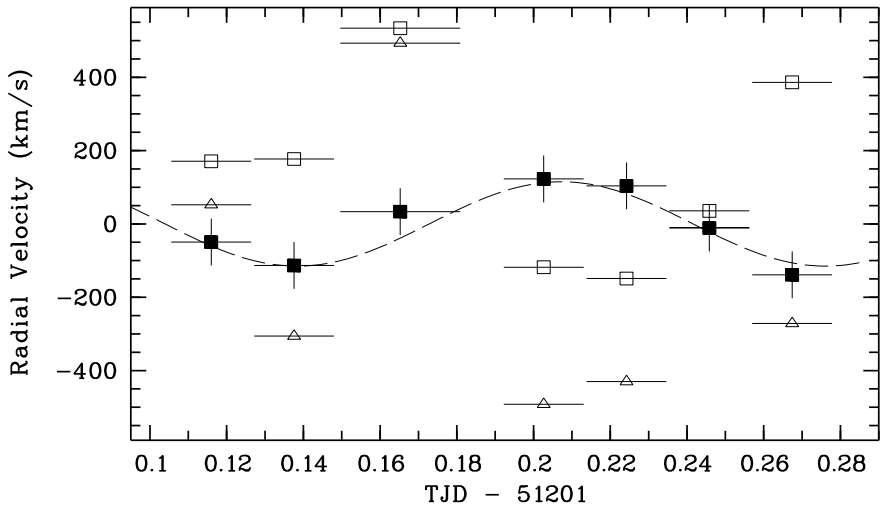,width=11.8cm,%
           bbllx=2.1cm,bblly=3.2cm,bburx=19.3cm,bbury=13.3cm,clip=}}\par
    \vspace*{-5.7cm}\hspace*{12.cm}
    \parbox[b]{55mm}{
    \caption[rvc]{Radial velocity curves of the \ion{He}{ii} $\lambda$4686 
       emission line (filled squares) and of the H$\beta$ (open squares) and
        H$\gamma$ (open triangle) absorption lines as measured from the 
        seven spectra taken in January 1999.
       The systemic velocity of 346 km/s has been subtracted. 
       The emission line data are fitted with a sinusoidal curve (dashed 
        line),  resulting in a best-fit period of 3.3 hours. 
       The velocity of the Balmer absorption lines seems not to be sinusoidal,
        and the amplitude is considerably larger than those of the \ion{He}{ii}
       line. 
             }
      \label{rv}
      }
   \end{figure*}

The system is very blue and has a mean magnitude and color of
$V$ = 19.7 mag and $B-V$ = --0.05.
In Fig. 1 we show the $V$ light curves of all individual nights.
The light curves of both 1998 nights and of January 26, 1999 are characterized
by an overall variation of $\sim$ 0.2 mag amplitude and rather smooth minima. 
In contrast, the system was much more variable on January 24 and 25, 1999 
(peak-to-peak amplitude $\sim 0.5$ mag) and the minima appeared more 
asymmetric or skewed. Also, minimum light was substantially reduced 
down to $\sim 20^{\rm m}$.

In order to derive an orbital period we carried out a period search using the 
analysis-of-variance method (Schwarzenberg-Czerny 1989).
The resulting periodogram (Fig. \ref{perio}) shows a variety of maxima
which are predominantly caused by the 1 day and 1 year sampling rates.
In addition, the strong night-to-night 
variations of the system prevent any clear signal to emerge.
The most probable periods (Fig. \ref{perio}) are 
0.147275 d, 0.172157 d, 0.206914 d and 0.258718 d.

There is a clear difference between the short and the long period 
possibilities: while for the above two shorter periods 
(0.147275 d and 0.172157 d) the folded light curve is basically sinusoidal, 
it is double humped for the two longer periods. It is important to note 
that in both cases the primary minima have different brightnesses
(differing by 0.2 mag).

For completeness we have also checked all other optical objects within the
X-ray error circle for possible variability, but found no other variable
star with an amplitude $>$0.2 mag down to 22 mag.

\section{Spectroscopy}

Low-dispersion spectroscopy has been performed on 1999 Jan. 22 with the EFOSC 
instrument at the 3.6m telescope on La Silla (ESO). A 300 grooves/mm grating
has been used, giving a dispersion of 2.1 \AA/pix and covering the
wavelength range 3800--8000 \AA. With the use of a 1\farcs5 slit
(adapted to the seeing during that night) the FWHM resolution is 19 \AA.
The slit was oriented along the parallactic angle always.
A total of seven consecutive spectra have been taken with exposure times
of 1800 sec each except for the third spectrum with 2700 sec. Due to the 
faintness of the object and the suspected short orbital period the length
of the exposures was a compromise between achieving a reasonable S/N ratio
and minimizing the phase coverage of each exposure. One-dimensional spectra
were extracted and processed using standard MIDAS routines. 
Wavelength calibration was done using He-Ar spectra taken before and after
the sequence of spectra. The standard star GD108 was used for flux-calibration.

\begin{table}[bh]
\caption{\ion{He}{ii} $\lambda$4686 velocities (after subtraction of
a systemic 346\,km/s) and equivalent widths.}
\begin{tabular}{rcc} 
\hline \noalign{\smallskip}
HJD mid-exposure   & Velocity (km/s) &  EW (\AA) \\
\noalign{\smallskip} \hline \noalign{\smallskip}
  2451201.11598~ & ~\,--49  & 4.8 \\
         .13757~ & --113  & 2.9:$\!\!\!$ \\
         .16525~ & ~~+33  & 4.4 \\
         .20265~ & +123  & 3.9 \\
         .22424~ & +104  & 2.9 \\
         .24582~ & ~\,--11  & 3.8 \\
         .26741~ & --139 & 3.4 \\
  \noalign{\smallskip} 
  \hline
  \end{tabular}
  \label{rvtab}
\end{table}

   \begin{figure}
      \vbox{\psfig{figure=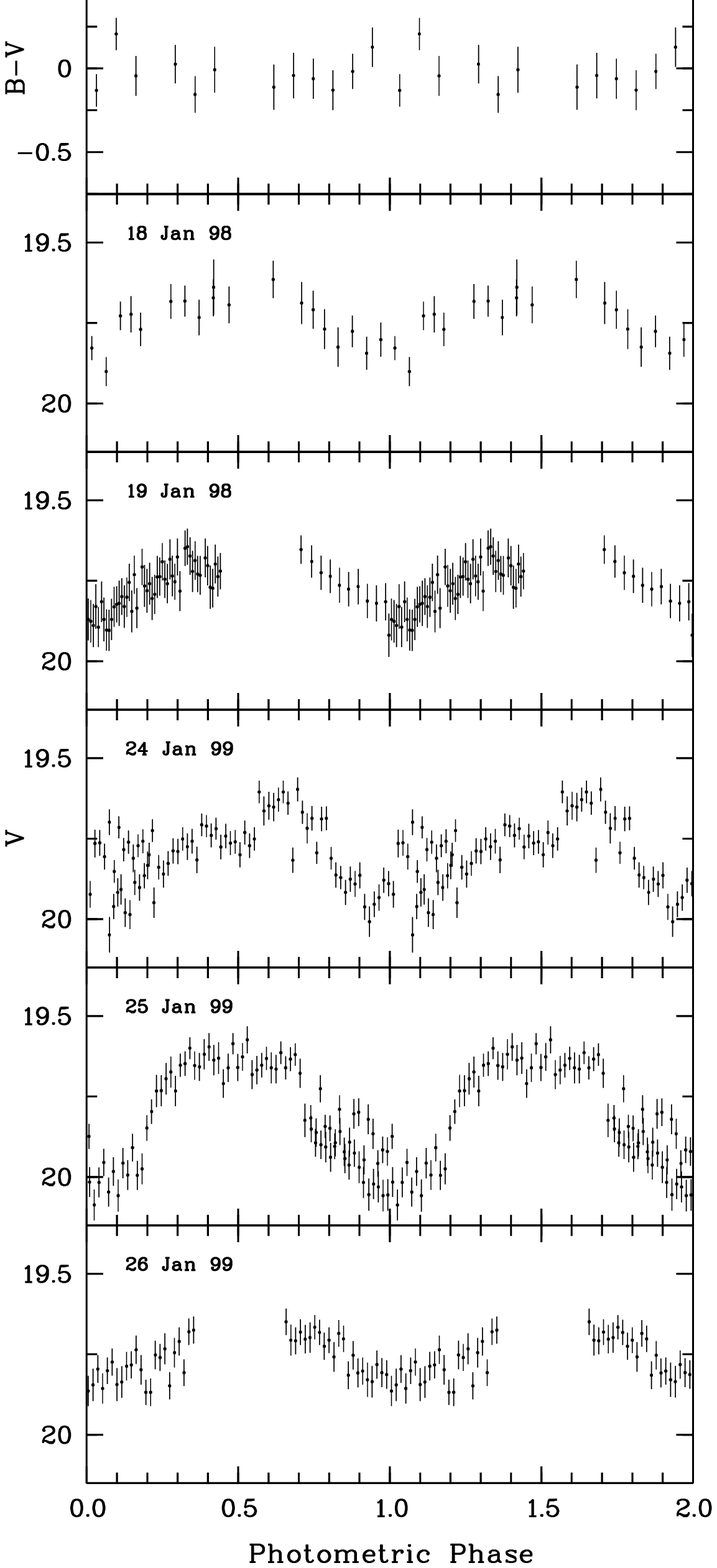,width=8.8cm,%
           bbllx=5.8cm,bblly=2.3cm,bburx=17.7cm,bbury=29.1cm,clip=}}
    \caption[foldd]{Photometric data folded with the adopted
            period of 0.147275 d. The top panel shows $B-V$. }
   \label{fold}
   \end{figure}

The spectrum of \rx0537\  has a steep, blue continuum and
a moderately strong \ion{He}{ii} $\lambda$4686 emission line.
The Balmer series appears in absorption except for H$\alpha$ which seemingly
is filled (Fig. \ref{ospec}). No other notable features
are seen in the individual spectra. The phase-averaged sum of all spectra
(Fig. \ref{ospec}) allows to identify a few other emission lines, among 
them \ion{He}{ii} $\lambda$4541, 
the \ion{C}{iii}/\ion{N}{iii} $\lambda\lambda$4630--4650 complex, 
and possibly also \ion{O}{vi} $\lambda$5290.

Despite the low S/N of the individual spectra 
we attempted a measurement of the \ion{He}{ii} $\lambda$4686 line 
strengths and velocities. The result is given in Tab. \ref{rvtab} and plotted 
in Fig. \ref{rv}. The \ion{He}{ii} $\lambda$4686 emission line shows a 
surprisingly clear radial velocity variation. Fitting a single sinusoidal 
curve yields a semiamplitude of $K$ = 115$\pm$20 km/s and a period of 
$P$ = 3.30$\pm$0.15 hrs ($\chi^2_{\rm red}$ = 0.95). 
The errors have been determined by a two-dimensional $\chi^2$ fitting,
i.e. the maximum and minimum possible periods have been determined
for various values of the phase zero-point.
If we use the standard procedure with a $\chi^2$-minimization only along
the y-axis (radial velocity), we get 
$K$ = 125$\pm$40 km/s and a period of $P$ = 3.45$\pm$0.25 hrs
($\chi^2_{\rm red}$ = 1.1).
The latter solution is plotted in the lower panel of Fig. \ref{perio}.
Though the errors for the spectroscopic period are large, 
the spectroscopic period matches one of our tentative photometric periods,
namely the shortest one at 0.147275 d. We therefore adopt this value
as the best orbital period, and the ephemeris for this is:
$$ T = (2451203.6392 \pm 0.0040 ) + (0.147275 \pm 0.0038) ~{\rm days}   $$

\begin{table*}
\caption{ROSAT X-ray observations of \rx0537}
\vspace{-0.2cm}
\begin{tabular}{llccccc}
\noalign{\smallskip}
\hline
\noalign{\smallskip}
  Date &  ROR & Detector & JD & count rate$^{(1)}$ (cts/s) & 
        N$_{cts}$ & HR1$^{(2)}$ \\
\noalign{\smallskip}
\hline
\noalign{\smallskip}
  900616      & 120006p   & PSPC & 2448058.5 & $<$0.0146 & & \\
  900618      & 110169p   & PSPC & 2448060.5 & $<$0.0121 & & \\
  900619      & 110175p   & PSPC & 2448061.5 & $<$0.0089 & & \\
  900619-23   & 110176p   & PSPC & 2448063.5 & $<$0.0079 & & \\
  900621      & 110182p   & PSPC & 2448063.5 & $<$0.0088 & & \\
  900710      & 120101p   & PSPC & 2448082.5 & $<$0.0131 & & \\
  911129-1202 & 400079p$^{(3)}$ & PSPC & 2448590.5 & $<$0.0153 & & \\
  920117-24   & 400079p$^{(3)}$ & PSPC & 2448642.5 & $<$0.0163 & & \\
  920509-16   & 300172p         & PSPC & 2448755.5 & 0.0200$\pm$0.0020 &
     92 & --0.93$\pm$0.09 \\
  921218-26   & 300172p-1       & PSPC & 2448978.5 & 0.0178$\pm$0.0037 &
     40 & --0.83$\pm$0.16 \\
  930615-27   & 300172p-2       & PSPC & 2449159.5 & $<$0.0049 & & \\
  931215-6    & 300335p         & PSPC & 2449338.0 & 0.0044$\pm$0.0008 & 
     45 & --1.00$\pm$0.15 \\
  940916      & 400352h         & HRI  & 2449611.5 & $<$0.0024 & & \\
\noalign{\smallskip}
\hline
\end{tabular}

\noindent{$^{(1)}$ Upper limits are 2$\sigma$ confidence, and for the
                   PSPC observations have been calculated using only 
                   the soft channels 11--50.\\
          $^{(2)}$ HR1 is the normalized count difference
    (N$_{\rm 52-201}$ -- N$_{\rm 11-41}$)/(N$_{\rm 11-41}$ + N$_{\rm 52-201}$),
                   where N$_{\rm a-b}$ denotes the number of counts in the 
                   PSPC between channel a  and channel b. \\
          $^{(3)}$ The investigation of this pointed observation has been 
                   split for the present paper into the two distinct
                   and widely separated temporal segments.
                      }
\label{xtab}
\end{table*}

The photometric data of all 5 nights, folded with this period, are shown
in Fig. \ref{fold}. This demonstrates more clearly that while the light
maxima have nearly identical brightnesses, the minima vary by slightly more
than 0.2 mag from one night to the other and to a somewhat lower degree 
even from one orbital cycle to the next (January 24 and 25, 1999). Also, the
irregular brightness fluctuations are clearly visible.

Due to the uncertainty of our final period and the 2 day difference between
spectroscopy and photometry we are unable to establish a relative phasing
between radial velocity and intensity modulation.

Attempts to measure the absorption line velocities of H$\beta$ and
H$\gamma$ are very difficult due to both, the poor S/N and the possible 
distortion by emission cores (H$\alpha$ is completely filled, and H$\delta$
is at the very edge of our wavelength coverage). 
Nevertheless, it is obvious that these
Balmer absorption lines exhibit a considerably larger velocity amplitude
than the \ion{He}{ii} emission line (Fig. \ref{rv}). Also, the velocity curve 
does not seem to be sinusoidal, and possibly out of phase with respect to
\ion{He}{ii} $\lambda$4686. A formal (one-dimensional) fit of a circular
orbit yields a period of 3.33 hrs, consistent within the errors to the 
\ion{He}{ii} emission line period.

   \begin{figure}
       \vbox{\psfig{figure=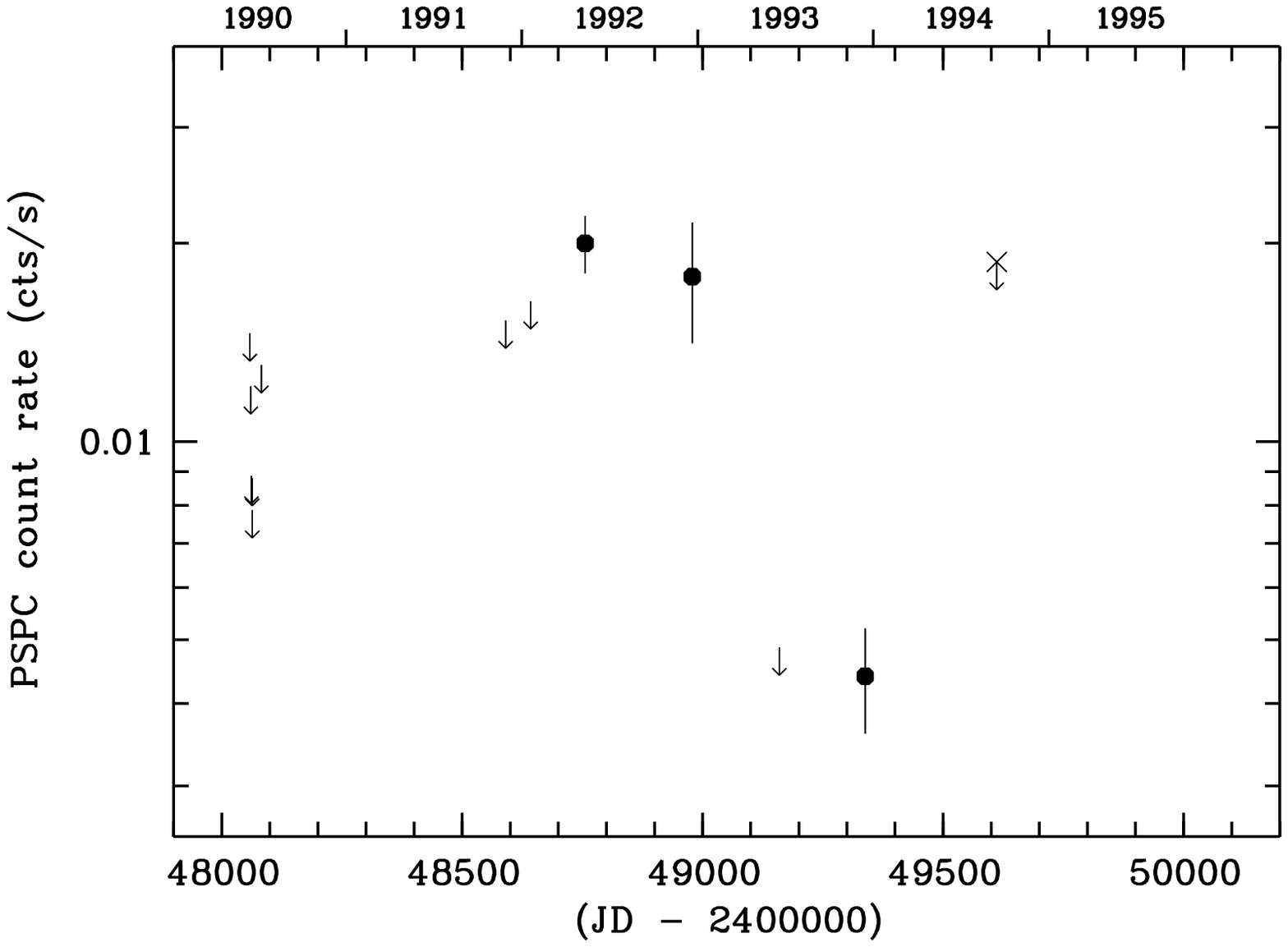,width=8.8cm,%
           bbllx=1.3cm,bblly=1.2cm,bburx=18.cm,bbury=13.2cm,clip=}}
    \caption[xlc]{X-ray light curve of \rx0537\ as observed with ROSAT.
           Arrows denote 2$\sigma$ upper limits, and the HRI upper limit
           on HJD = 2449611 has been transformed into a PSPC count rate
           upper limit using the conversion factor 7.8 (Greiner \etal\ 1996).
            }
   \label{xlc}
   \end{figure}

\section{Update of the X-ray light curve of \rx0537}

For the purpose of the discussion (see next paragraph) of alternative
interpretations it seems useful to have the complete ROSAT X-ray light curve
available. 
Though there has been no further ROSAT observation covering \rx0537\
since 1994 as presented in Orio \etal\ (1997),
we have recalculated the count rates or upper limits, respectively,
using the recently improved point spread function fitting and detection
algorithms within the EXSAS package
(Zimmermann \etal\ 1994). In addition, we used only PSPC channels 11--50
(corresponding to roughly 0.1--0.5 keV)
for the upper limit calculation since the source in its on-state has
such a very soft spectrum. Therefore, the intensity numbers of Tab. \ref{xtab}
differ slightly from those given by Orio \etal\ (1997). A visual representation
of the table data is given in Fig. \ref{xlc}.

\section{Discussion}
 
\subsection{Orbital variations and binary parameters}

There are basically two different models to explain the optical
orbital variations in supersoft X-ray binaries: 
(1) the optical light is produced by reprocessing of the white dwarf emission 
on the accretion disk, and the orbital variations are due to the varying 
aspect of the illuminated, heated, and flared accretion disk
(Schandl \etal\ 1997);
(2) these variations are due to the changing aspect of the irradiated
secondary (van Teeseling \etal\ 1998).
An irradiated secondary, if dominating the optical light, would produce
a smooth, sinusoidal light curve. This indeed is what we observe on three
out of the five nights. However, the irregular brightness variations
and minima of different depths as well as the asymmetric minima
during the two other nights are exactly what
the flared accretion disk would produce when the accretion rate varies
and the interaction of the accretion stream with the disk causes either
a varying amount of ``spray'', e.g. matter splashing at the stream-disk
impact site (Schandl \etal\ 1997) or a varying accretion disk rim 
height (Meyer-Hofmeister \etal\ 1997). 
Based on our available data we
cannot distinguish between these two alternatives.

The origin of the \ion{He}{ii} emission line is still a mystery.
If it came from the illuminated side of the donor, its radial velocity
amplitude would have to be 300--400 km/s, much higher than the observed
value of 115 km/s. Therefore, it is generally assumed that the 
\ion{He}{ii} emission line originates near to the accretor. Indeed,
the expected velocity amplitude of the accretor (for our best period)
is in the range 100--150 km/s.

\begin{figure}
      \vbox{\psfig{figure=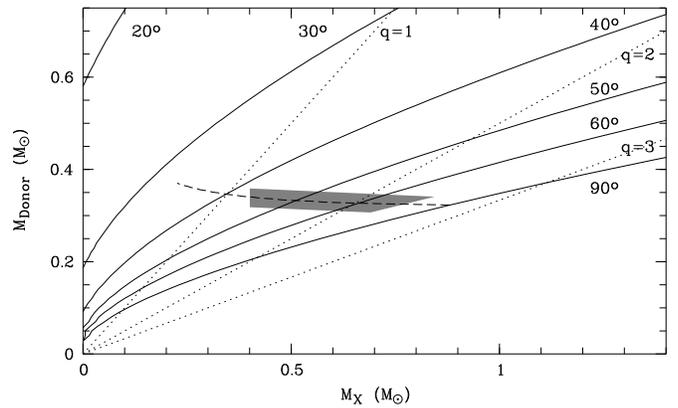,width=8.8cm,%
           bbllx=2.6cm,bblly=3.0cm,bburx=17.4cm,bbury=12.2cm,clip=}}
      \caption[mass]{Possible range of binary component masses based on
            our mass function of f(M)=0.023 \msun. Solid lines mark various 
            inclinations of the binary plane, and dotted lines
            visualize constant mass ratio $q$ = $M_{\rm X}/M_{\rm Donor}$.
            The dashed line indicates where the donor fills its Roche lobe
            (using the mass-radius relationship of Patterson 1984).
            The shaded area represents our best-estimate parameter space
            for the component masses (see text).
            }
      \label{massfct}
   \end{figure}

If we assume that the measured \ion{He}{ii} emission-line velocity is roughly
similar to the motion of the accretor, we may derive an estimate of the
donor mass. Combining the orbital period and the velocity amplitude results 
in a mass function of f(M) = 0.023 $\pm$ 0.014 \msun. If we assume the donor
to be a main-sequence star filling its Roche lobe, the resulting mass for
the donor is restricted to a narrow range of 0.32--0.37 \msun,
nearly independent of the inclination (Fig. \ref{massfct}).

The low mass of the donor implies that it does not contribute to the 
total light, since the apparent magnitude of a 0.35 \msun\ star at the 
LMC distance is $m_V \sim 29$ mag. Thus, the optical emission as well
as its orbital modulation must be caused differently, e.g. by the
(irradiated) accretion disk. Also, it is very unlikely that the 
Balmer absorption lines arise in the illuminated secondary, unless the
illumination increases the donor emission by 10 magnitudes.
Since the white dwarf itself is also an unlikely source, we therefore 
tentatively assign the Balmer absorption lines to either the accretion disk
or a possible wind. Substantially better data are needed to establish
the velocity variations of the Balmer absorption lines, and thus shed more
light on their nature.

The lack of detectable (at our accuracy) $B-V$ color variation through the 
orbital cycle of \rx0537\ suggests that the inner, bluer (hotter) part of 
the accretion disk are never occulted (during the 18 Jan. 1998 $B$ band
observation). This implies that the inclination is smaller than 
70\degr--74\degr.
Since \rx0537\ has shown luminous supersoft X-ray emission for the period
of approximately one year, the white dwarf mass should be large enough to allow
hydrogen burning. Previous investigations suggest a minimum mass of
0.4 \msun\ (Sienkewicz 1980).
This in turn implies 
(using Fig. \ref{massfct}) that the inclination should be $>$45 degrees.
We therefore tentatively assume $i \sim$ 45\degr--70\degr\ in the following.

Since the inclination cannot be larger than 90 degrees, Fig. \ref{massfct} also
implies that the white dwarf mass is certainly smaller than 0.9 \msun.
With the above best-estimate inclination range the mass of the white dwarf
is $<$ 0.8 \msun.

\subsection{On the nature of \rx0537}

We have determined a period of $\sim$3.5 hrs for the supersoft X-ray source
\rx0537. Since this is a period in the range typically populated by 
(low-luminosity) cataclysmic variables it is worth mentioning here that
the systemic velocity of \rx0537\ of $\sim$350 km/s together with the 
characteristic spectrum of a SSS leaves no doubt that \rx0537\ is a
supersoft X-ray source in the Large Magellanic Cloud, and not a galactic
foreground cataclysmic variable. The lack of any other variable object within 
the X-ray error circle together with the improved X-ray position and the 
good positional match clearly suggest that the optical object studied here 
is indeed the optical counterpart of \rx0537. 

Many properties of \rx0537\ are very similar to those of SMC 13, a supersoft
X-ray source in the Small Magellanic Cloud..
SMC 13 $\equiv$ 1E 0035.4--7230 $\equiv$ RX J0037.3--7214
(1RXS J003723.2--721415), has
a 4.1 hrs orbital period (Schmidtke \etal\ 1996, Crampton \etal\ 1997,
van Teeseling \etal\ 1998).
In SMC 13 the Balmer absorption line system  moves in phase with
the \ion{He}{ii}  emission lines, but with a larger amplitude. 
Based on the interpretation of the light curve as partially eclipsing 
(which constrains the 
inclination to 70\degr $< i <$ 78\degr) and the \ion{He}{ii} radial velocity
data, Crampton \etal\ (1997) derive a donor mass of 0.4--0.5 \msun\
and a mass of the accreting object of 1.3--1.5 \msun.
In contrast,  van Teeseling \etal\ (1998) interpret the
orbital modulation by the varying aspect of the illuminated secondary,
and derive an inclination of 20\degr $< i <$ 50\degr.
Also, fitting non-LTE models to the BeppoSAX X-ray spectrum,
Kahabka \etal\ (1999) derive a white dwarf mass of 0.6--0.7 \msun\,
consistent with the inclination range of van Teeseling \etal\ (1998).
Thus, \rx0537\ shares the following  characteristics with  SMC 13:
(i) the short orbital period, 
(ii) an almost identical optical spectrum,
(iii) the existence of Balmer absorption lines,
(iv) the different velocities of \ion{He}{ii} emission and Balmer 
absorption lines,
(v) the low visual absolute magnitude (see Tab. \ref{mvtab}). 
(vi)  the luminous, soft X-ray spectrum, and
(vii)  possibly a low white dwarf mass.
But there is also one clear
difference: \rx0537\ shows a factor of 7 X-ray variability over
the past 8 years while \1e\ has been completely constant 
(e.g. Kahabka \etal\ 1999).

With $P_{\rm orb} \sim 3.5$ hrs \rx0537\ is the shortest-period binary among 
the SSS. This implies that the standard scenario of SSS, in which the donor
is assumed to be more massive than the accreting white dwarf to ensure high
mass transfer rates on a thermal timescale (van den Heuvel \etal\ 1992), 
is not applicable for this system. Under the assumption of Roche-lobe
filling and a minimum white dwarf mass for steady-state burning of 
$\sim$0.5--0.6 \msun\ (Fujimoto 1982, Sion \& Starrfield 1994,
Cassisi \etal\ 1998), a mass ratio $ q \equiv M_{\rm Donor} / M_{\rm WD} > 1$ 
implies that $P_{\rm orb} \gax 4.5-5$ hours always. In addition, with a donor
mass of about 0.35 \msun\ it seems impossible to have a less massive white 
dwarf with H burning. Thus, \rx0537\ clearly does not fit 
this standard scenario.  

\begin{table}
\caption{Comparison of brightness and colours of SSS. We assume here
18.5 mag as the distance modulus for the LMC (Panagia et al. 1991),
18.8 mag for the SMC and reddening E(B-V)=0.065 and E(B-V)=0.043 for the
LMC and SMC, respectively.}
\begin{tabular}{lccccl} 
\hline \noalign{\smallskip}
   SSS   & $P$ (d) & $M_{\rm V}$  &  $B$--$V$  & $U$--$B$  & Refs.$^{(1)}$ \\
\noalign{\smallskip} \hline \noalign{\smallskip}
CAL 83           & 1.04 & --1.3 & --0.06 &       & 1, 2 \\
CAL 87           & 0.44 & +0.1 &   +0.1  &       & 3 \\
\rx0537\         & 0.14 & +1.0 & --0.03 & --0.69 & 4, 5 \\
SMC 13           & 0.17 & +1.4 & --0.13 & --1.13 & 6, 7, 8 \\
RX J0439.8--6809 & ~0.15? & +2.9 & --0.2 & --1.25 & 9, 10 \\
GQ Mus$^{(2)}$   & 0.06 & +2.7 & --0.3 & --1.1 & 11 \\
  \hline \noalign{\smallskip} 
  \end{tabular}

\noindent{\Ni\small $^{(1)}$ References: 
                    (1) Crampton \etal\ (1987),
                    (2) Smale \etal\ (1988),
                    (3) Schmidtke \etal\ (1993),
                    (4) Orio \etal\ (1997),
                    (5) this work,
                    (6) Schmidtke \etal\ (1996),
                    (7) Crampton \etal\ (1997),
                    (8) van Teeseling \etal\ (1998),
                    (9) Schmidtke \& Cowley (1996),
                   (10) Teeseling \etal\ (1996),
                   (11) Diaz \& Steiner (1989) \\
                    $^{(2)}$ During the phase as a supersoft X-ray source. }
  \label{mvtab}
  \end{table}

Similar arguments, though not as strong (because the component parameters
are less constrained), have been put forward for SMC 13 (Crampton \etal\ 1997).
Thus, there are now two sources which require a new donor scenario.
Alternative scenarios are either symbiotic
systems (see Kahabka \& van den Heuvel 1997), classical novae (e.g.
\"Ogelman et al. 1993), the systems that Kahabka and Ergma (1997)
define ``SMC 13'' type, and wind-driven binaries (van Teeseling \& King 1998).
Symbiotics, involving giant donors, have much too long orbital periods
to be applicable to \rx0537. A classical post-nova interpretation 
(similar to e.g. GQ Mus; \"Ogelman \etal\ 1993) is also unlikely
since the LMC was 
continuously monitored to search for novae in the last 50 years,
and a novae would have been detected in the optical passband.
Thus, there are two serious alternative explanations for \rx0537:
\begin{enumerate}
\item SMC 13 systems:
SMC 13 is  interpreted by Kahabka \& Ergma  (1997) as a cataclysmic variable
with a low mass white dwarf (0.6--0.7 M$_\odot$ white dwarf) having a
thick helium buffer layer. Such a system must have been a classical nova in
which, after a number of outbursts, the white dwarf  was heated up.
As a consequence the flashes have become very mild, without actual mass loss,
and the system is proposed to presently be in a phase of residual hydrogen 
burning after a mild shell flash. Van Teeseling \& King (1998) have
argued on statistical grounds that such a phase should last longer than 
100 yrs. In order for the residual burning to last
long enough (at least about 20 years since the {\it Einstein} discovery; 
Seward \& Mitchell 1981), a low CO abundance of the burning matter and
a hot, low-mass (0.6--0.7 \msun) white dwarf are required.
Sion \& Starrfield (1994) have demonstrated that hot white dwarfs with masses
as low as 0.5 \msun\ can stably burn hydrogen in a steady state.

\item wind-driven binaries: 
Van Teeseling \& King (1998) have shown that the strong
X-ray flux in supersoft sources should excite a strong wind 
($\dot M_{\rm wind} \sim 10^{-7}$ \msun/yr) from the irradiated companion
which in short-period binaries would be able to drive Roche lobe overflow 
at a rate comparable to $\dot M_{\rm wind}$. This may self-consistently
sustain stable recurrent or steady-state hydrogen burning on the white dwarf.
In binaries with a low-mass companion, the angular momentum loss by the wind
may dominate the binary evolution, and cause the period to increase with time.
One important open question of this scenario is how a system could
eventually enter such a self-consistent wind-driven phase.
Van Teeseling \& King (1998) suggested either a preceding SMC 13 phase,
or a late helium shell flash of the cooling white dwarf after the binary
has already come into contact as a cataclysmic variable.
\end{enumerate}

Because of the X-ray variability of \rx0537\ its secure classification
into one of the above two scenarios seems difficult.
One could expect that the SMC 13 systems should display a 
rather smooth X-ray light curve, unless a mild flash is occurring.
Such a flash should be accompanied by some optical brightening, however.
Also, if the X-ray turn-off of \rx0537\ were to be explained by an only short 
phase ($\sim$1 year as observed)  of  burning, the white dwarf mass 
must be high ($>$1.1 \msun), contrary to our findings (see Fig. \ref{massfct}).
Finally, the residual burning (between the H flashes) should be
connected to a temperature
increase (Sion \& Starrfield 1994), while the observations tend to
suggest the opposite behaviour (see the hardness ratio values in 
Tab. \ref{xtab} which are softest $\equiv$ coolest during the last
detection).

Whether or not the wind-driven supersoft source scenario is valid,
remains to be seen:
(i) The drop in X-ray intensity in 1993 is very probably not caused by
  increased absorption in the wind, otherwise the sensitive hardness 
 ratio HR1 would have changed substantially;
(ii) since it is difficult
to enter the self-excited wind phase, any repeated on-states would be
hard to explain -- assuming that the observed X-ray decline corresponds
to the sudden end of this wind-driven phase; 
(iii) this phase would
have lasted less than two years (Fig. \ref{xlc}) -- was it just an
unsuccessful attempt to enter the wind-driven phase?
Overall, the  wind-driven supersoft source scenario seems to provide 
a possible explanation for \rx0537, though it remains unclear what the 
observational consequences would be.

\section{Conclusions}

We have obtained new photometric and spectroscopic data of the supersoft
X-ray source \rx0537\ which show that
\begin{enumerate}
\vspace{-0.2cm}\item the orbital period is $\sim$3.5 hrs,
\item the inclination of the orbital plane is 45\degr\ $\lax i \lax$ 70\degr,
\item the component masses are $0.4 \lax M_{\rm accretor} \lax 0.8$ \msun\ and
     $0.32 \lax M_{\rm donor} \lax 0.37$ \msun, respectively, 
\item the canonical scenario of thermal-timescale mass transfer
  from a donor more massive than the white dwarf (van den Heuvel \etal\ 1992)
  is clearly not applicable,
\item the binary system \rx0537\ could be a SMC 13 system or wind-driven
  system, though the observed X-ray variability poses problems for both
  scenarios. 
\vspace{-0.2cm}
\end{enumerate}

\rx0537\ is still quite puzzling in many respects, and 
 new observations at  different wavelengths should help solving the
 riddle. The important and exciting fact is that it is a new type of
 transient X-ray source, possibly representative of a whole new
 class, and understanding it will also add to a  better understanding of
 close binary evolution.

\begin{acknowledgements}
JG and RS are supported by the German Bundes\-mini\-sterium f\"ur Bildung,
Wissenschaft, Forschung und Technologie
(BMBF/DLR) under contract Nos. FKZ 50 QQ 9602 3 and 50 OR 9708 6,
and MO by the Italian Space Agency Research Program.

\end{acknowledgements}


\begin{thebibliography}{}


\bibitem[]{cit98} Cassisi S., Iben I.Jr., Tornambe A., 1998, ApJ 496, 376


\bibitem{cch87} Crampton D., Cowley A.P., Hutchings J.B., 
 \etal\ 1987, ApJ 321, 745

\bibitem{chc97} Crampton D., Hutchings J.B., Cowley A.P., Schmidtke P.C.,
   1997, ApJ 489, 903

\bibitem{ds89} Diaz M.P., Steiner J.E., 1989, ApJ 339, L41

\bibitem{fu82} Fujimoto M.Y., 1982, ApJ 257, 767

\bibitem{gre96} Greiner J. (Ed.), 1996, Supersoft X-ray Sources,
     Lect. Notes in Physics 472, Springer 

\bibitem[]{gsh96} Greiner J., Schwarz R., Hasinger G., Orio M., 1996, 
      A\&A 312, 88

\bibitem{ke97} Kahabka P., Ergma E., 1997, A\&A 318, 108

\bibitem{kvdh97} Kahabka P., van den Heuvel E.P.J., 1997, ARAA 35, 69

\bibitem{kph99} Kahabka P., Parmar A.N., Hartmann H.W., 1999, A\&A 346, 453

\bibitem[]{lan92} Landoldt A.U., 1992, AJ 104, 340

\bibitem{lhg81} Long K.S., Helfand D.J., Grabelsky D.A., 1981, ApJ 248, 925

\bibitem[]{ms89} Mateo M., Schechter P., 1989, in 
                {\it 1st ESO/ST-ECF Data Analysis Workshop}
                eds. P.J. Grosbol, F. Murtagh \& R.H. Warmels, p. 69

\bibitem[]{msm97} Meyer-Hofmeister E., Schandl S., Meyer F., 1997, A\&A 321, 
    245

\bibitem{motch93} Motch C., Werner K., Pakull M.W., 1993, A\&A 268, 561

\bibitem{ooks93} \"Ogelman H.,  Orio M., Krautter J., Starrfield S., 1993,
   Nat. 361, 331

\bibitem{oo93} Orio M., \"Ogelman H., 1993, A\&A 273, L56

\bibitem{oo95} Orio M., 1995, in ``Cataclysmic Variables'', eds. A. Bianchini,
   M. Della Valle and M. Orio, Dordrecht Boston: Kluwer
  Academic Publishers, p. 429

\bibitem{ovm97} Orio M., Della Valle M., Massone G., \"Ogelman H., 1997, 
   A\&A 325, L1

\bibitem{pba87} Pakull M.W., Beuermann K., Angebault L.P., Bianchi L., 1987,
   Ap\&SS 131, 689

\bibitem{pbkp88} Pakull M.W., Beuermann K., van der Klis M., van Paradijs J., 
   1988, A\&A 203, L27

\bibitem{pa91} Panagia N., Gilmozzi R., Macchetto F., Adorf H.-M.,
   Kirshner R.P., 1991, ApJ 380, L23

\bibitem[]{patt84} Patterson J., 1984, ApJS 54, 443

\bibitem{rds94} Rappaport S., DiStefano R., Smith J.D., 1994, ApJ 426, 692

\bibitem{smm97} Schandl S., Meyer-Hofmeister E., Meyer F., 1997, A\&A 318, 73

\bibitem{smc93} Schmidtke P.C., McGrath T.K., Cowley A.P., Frattare L.M., 1993,
   PASP 105, 863

\bibitem{sc96} Schmidtke P.C., Cowley A.P., 1996, AJ 112, 167

\bibitem{scg96} Schmidtke P.C., Cowley A.P., Mc Grath T.K., Hutchings
  J.B., Crampton D.,  1996, AJ 111, 788

\bibitem[]{s89} Schwarzenberg-Czerny A., 1989, MNRAS 241, 153

\bibitem[]{sm81} Seward F.D., Mitchell M., 1981, ApJ 243, 736

\bibitem[]{s80} Sienkewicz R., 1980, A\&A 85, 295

\bibitem[]{ss94} Sion E.M., Starrfield S.G., 1994, ApJ 421, 261

\bibitem[]{scc88} Smale A.P., Corbet R.H., Charles P.Q., \etal,
   1988, MNRAS 233, 51

\bibitem{vdh92} van den Heuvel E.P.J., Bhattacharya D., Nomoto K., Rappaport
   S.A., 1992, A\&A 262, 97 

\bibitem{vtk98} van Teeseling A., King A.R., 1998, A\&A 338, 957

\bibitem{trb96} van Teeseling A., Reinsch K., Beuermann K., 1996, A\&A 307, L49

\bibitem{trh97} van Teeseling A., Reinsch K., Hessman F.V., Beuermann K., 1997,
   A\&A 323, L41

\bibitem{trpb98} van Teeseling A., Reinsch K., Pakull M.W., Beuermann K., 1998,
  A\&A 338, 947

\bibitem{wan91} Wang Q., 1991, MNRAS 252, 47p

\bibitem[]{zbbd94} Zimmermann H.U., Becker W., Belloni T., D\"obereiner S., 
          Izzo C., Kahabka P., Schwentker O.,  1994, MPE report 257

\end{thebibliography}
\end{document}